\documentstyle[12pt]{article}

\begin{document}
\topmargin 0pt
\oddsidemargin 7mm
\headheight 0pt
\topskip 0mm

\addtolength{\baselineskip}{0.40\baselineskip}

\hfill SNUTP 94-32

\hfill SOGANG-HEP 189/94

\hfill March 30, 1994

\begin{center}

\vspace{36pt}
{\large \bf Batalin-Tyutin Quantization of the (2+1) dimensional
        nonabelian Chern-Simons field theory}

\end{center}

\vspace{36pt}

\begin{center}

Won Tae Kim$^*$

\vspace{20pt}
{\it Center for Theoretical Physics and Department of Physics, \\
Seoul National University, Seoul 151-742, Korea}

\vspace{12pt}

Young-Jai Park$^\dagger$

{\it Department of Physics and Basic Science Research Institute, \\
Sogang University, C.P.O. Box 1142, Seoul 100-611, Korea}

\end{center}

\vfill

\begin{center}
{\bf ABSTRACT}
\end{center}
The (2+1) dimensional nonabelian Chern-Simons theory coupled to complex
scalar fields is quantized by using the Batalin-Tyutin canonical Hamiltonian
method which systematically embeds second-class constraint system into
first-class one.
We obtain the gauge-invariant nonabelian Wess-Zumino type action in
the extended phase space.

\vspace{12pt}

\noindent

\vspace{24pt}

\vfill
\vspace{2cm}
\hrule
\vspace{0.5cm}
\hspace{-0.6cm}$^*$ E-mail address : wtkim@phya.snu.ac.kr \\
$^\dagger$ E-mail address : yjpark@ccs.sogang.ac.kr

\newpage

To quantize the second-class constraint system, which does not form a closed
constraint algebra in Poisson brackets, the Dirac method has been widely used
in the
Hamiltonian formalism [1].  The brackets (commutators) compatible with
constraints and the dynamical equations of motion in the second-class system
are evaluated without choosing gauge conditions or additional constraints
since the Dirac matrices are invertible.
In general, since the resulting brackets are field-dependent and nonlocal,
and have a serious ordering problems
between field operators, these are under unfavorable circumstances
in finding canonically conjugate pairs.

On the other hand,
quantizations of the first-class  constraint systems [2,3] have been well
appreciated in a gauge invariant manner, i.e., preserving
Becci-Rouet-Stora-Tyutin (BRST) symmetry [4,5].  If the second-class constraint
system can be converted into first-class one in an extended phase space,
Dirac brackets can not be needed and then the remaining quantization
program is to follow the method of Ref. [2-5].
This procedure has been extensively studied by Batalin, Fradkin, and Tyutin
[6,7] in the canonical formalism. In the path-integral framework, especially
concentrating
on the second-class gauge algebra, which is related to the gauge anomaly,
Faddeev and Shatashivili [8] introduced the Wess-Zumino action [9], which
cancels the gauge anomaly and gives the first-class  gauge algebra. After their
work, in the
Hamiltonian method, the Wess-Zumino actions for the various models [10,11,12]
have been studied following Refs. [6,7].

Recently, Banerjee [13] has pointed out an interesting application of
Batalin-Tyutin Hamiltonian method [7]
to the second-class constraint system of
the abelian Chern-Simons (CS) field theory [14-16], which yields a strongly
involutive constraint algebra in an extended phase space,
and obtained a new abelian Wess-Zumino
type action, which cannot be derived in the usual path-integral framework.
As shown in his work,
the nature of second-class constraint algebra originates from the
symplectic structure of CS term, not due to the local gauge symmetry
breaking.  There are some other interesting examples in this direction [17].

In this paper, we shall apply the Batalin-Tyutin Hamiltonian method [7]
to the nonabelian CS field theory by using the recent
progress of nonabelian Batalin-Tyutin quantization [18], which involves the
weakly involutive constraint algebra.
The phase space partition function is constructed in order to connect with the
Lagrangian formulation, and explicitly evaluated in two special gauges, unitary
and Faddeev-Popov gauges [19].
As a result, the nonabelian
Wess-Zumino like action is obtained in a local gauge invariant fashion in the
extended phase space by finding the suitable gauge transformation of
new dynamical fields.

Let us now consider the (2+1) dimensional nonabelian CS theory coupled to the
nonabelian scalar fields whose dynamics is given by
\begin{equation}
  S_0 = \int  d^3 x~ \left(   \kappa \epsilon^{\mu\nu\rho}
                             tr( A_{\mu}\partial_{\nu}A_{\rho}
                           - \frac{2}{3}iA_{\mu}A_{\nu}A_{\rho} )
                       + (D_{\mu}\phi)^{\dagger} (D^{\mu}\phi) \right),
\end{equation}
where $\phi$ is an N-component scalar field, which transforms according to the
fundamental representation of the gauge group, and the covariant derivative is
$D_{\mu} = \partial_{\mu} - iA_{\mu}$ (the matrix-valued gauge field
$A_{\mu}=A_{\mu}^aT^a$).
We adopt the conventions, $g_{\mu\nu}={\rm diag} (+,-,-)$, and
$\epsilon^{012}=+1$.
The matrices $T^a$'s in the fundamental representation are Hermitian group
generators
satisfying $[T^a,T^b]=if^{abc}T^c$, and are normalized as
$tr(T^aT^b)=\frac{1}{2}\delta_{ab}$.
Note that the gauge invariance of the nonabelian CS term requires
the quantization of the dimensionless constant $\kappa$, $\kappa=\frac{n}{4\pi}
(n \in Z)$ [14].

The canonical momenta of gauge fields and scalar fields are given by
\begin{eqnarray}
  \Pi_0^a &\approx& 0,  \nonumber  \\
  \Pi_i^a &=& \frac{\kappa}{2}\epsilon_{ij}A^{ja}~~~(i = 1, 2),  \nonumber  \\
  \Pi_{\phi} &=& (D_0\phi)^{\dagger},  \nonumber \\
  \Pi_{\phi}^{\dagger} &=& (D_0\phi).
\end{eqnarray}
The $\Pi_0^a$ and $\Pi_i^a$ are primary constraints [1].  The canonical
Hamiltonian is
\begin{equation}
  H_c = \int d^2 {\bf x}~ \left( \Pi_{\phi}\Pi_{\phi}^{\dagger}
                     + (D_i\phi)^{\dagger}(D_i\phi) - A_0^aG^a \right),
\end{equation}
where $G^a$ is the Gauss' law constraint defined by
\begin{equation}
  G^a = \frac{\kappa}{2}\epsilon^{ij}F_{ij}^a + J_0^a.
\end{equation}
Here, the field strength is $F_{ij}^a = \partial_iA_j^a-\partial_jA_i^a
+ f^{abc}A_i^bA_j^c$, and the charge density is $J_0^a = -i(\Pi_{\phi}T^a\phi
- \phi^{\dagger}T^a\Pi_{\phi}^{\dagger})$.  The time evolution of the Gauss'
law constraint (4) generates no more additional constraints.
To obtain the maximally irreducible
constraint system, we redefine the above primary and secondary constraints
similarly to the abelian case in
Refs. [16,20] as follows
\begin{eqnarray}
  \Omega_0^a &=& \Pi_0^a \approx 0,  \\
  \Omega^a &=& (D_i\Pi^i)^a + J_0^a
             + \frac{\kappa}{2}\epsilon^{ij}\partial_iA_j^a \approx 0,  \\
  \Omega_i^a &=& \Pi_i^a - \frac{\kappa}{2}\epsilon_{ij}A^{ja} \approx 0.
\end{eqnarray}
Then the first-class constraint algebras by using the Poisson brackets are
\begin{eqnarray}
  \{ \Omega_0^a(x), \Omega_i^b(y) \} &=&
                   \{ \Omega_0^a(x),\Omega^b(y) \} = 0,  \nonumber \\
  \{ \Omega^a(x), \Omega^b(y) \} &=&
                   f^{abc}\Omega^c(x)\delta^2({\bf x}  - {\bf y} ),  \nonumber
\\
  \{ \Omega_i^a(x), \Omega^b(y) \} &=&
                   f^{abc}\Omega_i^c(x)\delta^2({\bf x}  - {\bf y} ),
\end{eqnarray}
and the second-class constraint algebra is given by
\begin{equation}
\{ \Omega_i^a(x), \Omega_j^b(y) \} \equiv  \triangle^{ab}_{ij}(x, y)
                 =  - \kappa\epsilon_{ij}\delta_{ab}\delta^2({\bf x}  - {\bf y}
),
\end{equation}
where we denote $x=(t, {\bf x})$ and the two-space vector ${\bf x}=(x^1, x^2)$.
Therefore, $\Omega_0^a$ and $\Omega^a$ form first-class constraint system,
and we defer them.

Following to the Batalin-Tyutin quantization method [7,18],
let us now introduce new nonabelian auxiliary fields
$\Phi^{ia}$ to convert the second-class
constraint $\Omega_i^a$ into first-class one
in the extended phase space, and
assume the following Poisson algebra satisfying
\begin{equation}
   \{ \Phi^{ia}(x), \Phi^{jb}(y) \} = W_{ab}^{ij}(x,y).
\end{equation}
The modified constraint  $\tilde{\Omega}_i^a$ in the extended phase space is
given by
\begin{equation}
  \tilde{\Omega}_i^a(\Pi_i^a,A_i^a,\Phi^{ia})
         =  \Omega_i^a + \sum_{n=1}^{\infty}\Omega_i^{a(n)}
                       ~~~(\Omega_i^{a(n)} \sim (\Phi^{ia})^n )
\end{equation}
satisfying the boundary condition, $\tilde{\Omega}_i^a(\Pi_i^a,A_i^a,0)
= \Omega_i^a$.  The first order correction in the infinite series is
\begin{equation}
  \Omega_i^{a(1)}(x) = \int d^2 {\bf y} X_{ij}(x,y)\Phi^{j a}(y),
\end{equation}
and the first-class  constraint algebra of $\tilde{\Omega}_i^a$ requires the
condition as follows,
\begin{equation}
   \triangle_{ij}^{ab}(x,y) +
   \int d^2 {\bf w}~ d^2 {\bf z}~
        X_{ik}^{ac}(x,w) W_{cd}^{kl}(w,z)X_{lj}^{db}(z,y)
         = 0.
\end{equation}
As was emphasized in Ref. [13,17], there is a natural arbitrariness in choosing
$W_{ab}^{ij}$ and $X_{ij}^{ab}$ from Eq. (13), which corresponds to canonical
transformation in the extended phase space [6,7].  We take the simple solutions
as
\begin{eqnarray}
  W_{ab}^{ij}(x,y)
         &=& \epsilon^{ij}\delta_{ab}\delta^2({\bf x} -{\bf y}), \nonumber  \\
  X_{ij}^{ab}(x,y)
         &=& -\sqrt{\kappa}g_{ij}\delta_{ab}\delta^2({\bf x} -{\bf y}),
\end{eqnarray}
which are compatible with Eq. (13) as it should be.
Note that the above choices remarkably simplify the constraint algebra
and give the compact involutive Hamiltonian.
Then, the modified constraint
$\tilde{\Omega}_i^a$ gives a strongly first-class  constraint algebra,
\begin{equation}
  \{\tilde{\Omega}_i^a(x), \tilde{\Omega}_j^b(y) \} = 0,
\end{equation}
where $\tilde{\Omega}_i^a=\Omega_i^a - \sqrt{\kappa}~g_{ij}\Phi^{ja}$.
However, this is not a whole story for the constraint algebra.  The modified
constraint
$\tilde{\Omega}_i^a$ is still second-class constraint
because it does not form a first-class  algebra
with $\Omega^a$ as easily seen from Eq. (8).
Therefore, in order to obtain the fully first-class  constraint algebra
( weakly first class in the nonabelian structure ), we should also modify
$\Omega^a$ as
\begin{equation}
  \tilde{\Omega}^a = \Omega^a + Y^a,
\end{equation}
where we assume that $Y^a$ is only a function of new fields, $\Phi^{ia}$
and require two
conditions to maintain the weakly first-class  constraint structure as follows,
\begin{eqnarray}
  \{ Y^a(x), Y^b(y) \} &=&
                   f^{abc}Y^c(x)\delta^2({\bf x}  - {\bf y} ),  \nonumber  \\
  \{ \Phi^{ia}(x), Y^b(y) \} &=&
                   f^{abc}\Phi^{ic}(x)\delta^2({\bf x}  - {\bf y} ),
\end{eqnarray}
with the consistent choice of $Y^a$ as
\begin{equation}
  Y^a(x) = \frac{1}{2}\epsilon_{ij}f^{abc}\Phi^{ib}\Phi^{jc}.
\end{equation}
Then, one can easily check that the modified $\tilde {\Omega}_i^a$ and
$\tilde {\Omega}^a$ form the fully
first-class  constraint algebra,
\begin{eqnarray}
  \{ \Omega_0^a(x), \tilde{\Omega}_i^b(y) \} &=&
                   \{ \Omega_0^a(x), \tilde{\Omega}^b (y) \} = 0,  \nonumber \\
  \{ \tilde{\Omega}^a(x), \tilde{\Omega}^b(y) \} &=&
        f^{abc}\tilde{\Omega}^c(x)\delta^2({\bf x}  - {\bf y} ),  \nonumber  \\
  \{ \tilde{\Omega}_i^a(x), \tilde{\Omega}_j^b(y) \} &=&  0,  \nonumber  \\
  \{ \tilde{\Omega}_i^a(x), \tilde{\Omega}^b(y) \} &=&
                   f^{abc}\tilde{\Omega}^c_i(x)\delta^2({\bf x}  - {\bf y} )
\end{eqnarray}
by using Eqs. (16) and (18).

Let us now work out the involutive Hamiltonian [7] in the extended phase space
which
is similar to the abelian case [13]
except the nonabelian effect due to the
extended gauge group structure.  It is given by the infinite series [7],
\begin{equation}
  \tilde{H} = H_c + \sum_{n=1}^{\infty}H^{(n)}.
\end{equation}
Using the solution for the involution of $\tilde{H}$ for the nonabelian
case [18], it is given by
\begin{eqnarray}
  H^{(n)} &=& -\frac{1}{n} \int d^2 {\bf x} d^2 {\bf y} d^2 {\bf z}~
              \Phi^{ia}(x)W_{ij}^{ab}(x,y)X_{bc}^{jk}(y,z)G_k^{c(n-1)}(z)
              \nonumber  \\
          &=& \frac{1}{n\sqrt{\kappa}} \int d^2 {\bf x}~
              \epsilon^{ij}\Phi^{ia}(x)G_j^{a(n-1)}(x)  ~~~(n \geq 1),
\end{eqnarray}
where $W_{ij}^{ab}$ and $X_{ab}^{ij}$ are the inverse matrices of $W_{ab}^{ij}$
and $X_{ij}^{ab}$ respectively, and Eq. (14) is used in the second line in
Eq. (21).  For our case, the generating functions $G_k^{(n)}$ are given by
\begin{eqnarray}
  G_i ^{a(0)} &=& \{ \Omega_i^{a(0)}, H_c \}
                    = J_i^a - \kappa \epsilon^{ij}(D_j A_0)^a, \nonumber  \\
  G_i ^{a(1)} &=& \{ \Omega_i^{a(0)}, H^{(1)} \}_{\cal O}
                    + \{ \Omega_i^{a(1)}, H_c \}_{\cal O}  \nonumber  \\
             &=& -\frac{1}{\sqrt \kappa} \epsilon_{ij}\phi^{\dagger}
                 (T^a T^b + T^b T^a) \Phi^{jb} + {\sqrt \kappa} g_{ij} f^{abc}
                 \Phi^{jb} A_0^c,  \nonumber  \\
  G_i ^{a(n)} &=& 0 ~~~ (n \geq 2),
\end{eqnarray}
where the symbol ${\cal O}$ in Eq. (22) represents
that the Poisson brackets are
calculated among the original variables, i.e.,
${\cal O}=(\Pi_i^a,A^a)$.  Then, after
the $n=2$ finite truncations, we obtain the final
expression for the Hamiltonian ${\tilde{H}}$ as follows
\begin{eqnarray}
\tilde{H} = H_c &+& \int d^2{\bf x}~
            \left( - A_0^a ( \frac{1}{2}\epsilon_{ij}f^{abc}
              \Phi^{ib}\Phi^{jc} + \sqrt{\kappa}(D_i\Phi^i)^a ) \right.
              \nonumber  \\
  &-& \left. \frac{1}{\sqrt{\kappa}}\epsilon_{ij}\Phi^{ia}J^{ja}
            - \frac{1}{2\kappa}g_{ij}\Phi^{ia}\Phi^{jb}\phi^{\dagger}
                 (T^aT^b+T^bT^a) \phi \right).
\end{eqnarray}
Note that the compact form of involutive Hamiltonian (23) is a result of
the symmetric choice of $X_{ij}^{ab}$ (14) which is compared to the
abelian case [13].
By using the Poisson brackets, the following involutive relations
are given,
\begin{eqnarray}
  \{ \Omega_0^{a}~,\tilde{H} \} &=& \tilde{\Omega}^a
                            - (D_i\tilde{\Omega}^i)^a,  \nonumber  \\
  \{ \tilde{\Omega}^a~ ,\tilde{H} \}
                        &=&- f^{abc}A_0^b\tilde{\Omega}^c, \nonumber  \\
  \{ \tilde{\Omega}_i^a~ ,\tilde{H} \}
                        &=&  0.
\end{eqnarray}
In the above constraint analysis, from Eqs. (19) and (24),
we see that the original second-class
constraint system is converted into the first-class  system
if one introduces two fields,
which are conjugate each other in the extended phase space.
The origin of second-class constraint is due to the symplectic structure
as the abelian case [13], and the modifications of primary constraints
$\Omega^a_i$ are similarly done. However, since the set
$(\Omega_0^a, \Omega^a, \tilde{\Omega}^a_i)$ does not become
first-class  constraints,
the modification of $\Omega^a$ should be also needed to make the fully
first-class  constraint system (19).

Next we consider the partition function of the model
in order to present the Lagrangian corresponding to $\tilde{H}$
in the (canonical) Hamiltonian formalism. We relabel the constraints
as $\Gamma^a_0=\Omega_0^a, \Gamma^a_i=\tilde{\Omega}^a_i$,
and $\Gamma^a_3=\tilde{\Omega}^a$.  Then, the phase space partition function
is given by
\begin{eqnarray}
Z &=& \int {\cal D} \Pi^{\mu a} {\cal D} A^{\mu a}
           {\cal D} \Pi^{\dagger}_{\phi}
           {\cal D} \phi^{\dagger} {\cal D} \Pi_{\phi}
           {\cal D} \phi {\cal D} \Phi^{ia}
           \prod^3_{\alpha, \beta = 0}
            \delta(\Gamma^a_{\alpha}) \delta(G^b_{\beta})
            \mid det \{\Gamma^a_{\alpha}, G^b_{\beta}\} \mid
            e^{iS}, \nonumber \\
S &=& \int  d^3 x ~( \Pi^{\mu a} {\dot A_{\mu}^a} +
            \Pi^{\dagger}_{\phi} \dot{\phi} + \Pi_{\phi} \dot{\phi}
            + \Phi^{2a} {\dot \Phi^{1a}} )
            - \int dt~ \tilde {H},
\end{eqnarray}
where we regard new fields $( \Phi^{1a}, \Phi^{2a} )$
as conjugate pairs [13]
satisfying the Poisson algebra (10) and Eq. (14) , and $G^b_{\beta}$
are gauge fixing conditions to make the nonvanishing determinant of
$\Gamma^a_{\alpha}$ and $G^b_{\beta}$. The integration with respect to
the momenta $\Pi^{\mu a},\Pi^{\dagger}_{\phi}$ and $\Pi_{\phi}$,
and the exponentiation of the $\delta(\Gamma^a_3)$
by the Fourier transformation with variables $\xi^a$,
and the change of variables
as $A^a_0 \rightarrow A^a_0+\xi^a$ give the compact local action as follows
\begin{eqnarray}
      S &=& S_0+S_{\rm{WZ}}, \nonumber \\
      S_{\rm WZ} &=& \int d^3x~ \left(- \epsilon_{ij}
                   tr({\Phi}^i \partial_0 \Phi^j )-2\sqrt{\kappa} tr(\Phi^i
F_{i0})-2i
                  \epsilon_{ij} tr(A_0\Phi^i\Phi^j) \right. \nonumber \\
            &+& \left. \frac{2}{\sqrt{\kappa}}\epsilon_{ij}tr(\Phi^i J^j)
                 +\frac{1}{\kappa} \phi^{\dagger} (\Phi^i\Phi_i) \phi \right)
\end{eqnarray}
with the corresponding measure
\begin{equation}
[{\cal D} \mu] \equiv {\cal D} A^{\mu a} {\cal D}
                       \phi^{\dagger} {\cal D}\phi {\cal D} \Phi^{ia} {\cal D}
                      \xi^{a}
                      \prod^3_{\beta = 0}
                         \delta(G^a_{\alpha}[A^a_0 + \xi^a ,
                                          A^a_i , \Phi^{ia}])
                            \mid det \{\Gamma^a_{\alpha}, G^b_{\beta}\} \mid
\end{equation}
where $ J^a_i = -i ( (D_i \phi)^{\dagger} T^a \phi -  \phi^{\dagger}
T^a (D_i \phi)) $.
Note that as expected, if we choose the unitary gauge, i.e., $ \Phi^{ia} = 0 $,
the original theory is recovered.

It seems appropriate to comment that
the gauge invariance of $S_{\rm WZ}$ should be maintained
because the second-class constraint structure only comes from
the symplectic structure of Chern-Simons term.
It is implemented by the following ${\it extended}$ gauge transformations
involving
the transformation of the new fields,
\begin{eqnarray}
A^{'}_{\mu} &=& U A_{\mu} U^{-1} + i U \partial_{\mu}U^{-1}, \nonumber \\
\phi^{'} &=& U \phi, \nonumber \\
\phi^{\dagger '} &=& \phi^{\dagger} U^{-1}, \nonumber \\
\Phi^{i'} &=& U\Phi^i U^{-1},
\end{eqnarray}
where $\Phi^i={\Phi^i}^aT^a$,
and the gauge function $U=e^{i\theta^aT^a}$.
Note that the Wess-Zumino action in Eq. (26) is gauge invariant
in spite of the lack of the manifest Lorentz invariance.
On the other hand, since the unitary gauge recovered
the manifestly Lorentz invariant original action,
following the Fradkin-Vilkovisky theorem [2,3],
the actual invariance is maintained from the fact that the final result
for the partition function $Z$ is independent of the gauge fixing conditions.

For the Faddeev-Popov gauge [19] which does not involve the momenta in
the gauge fixing conditions,
the following nontrivial form of $S_{\rm WZ}$ is given by integrating
out the momentum,
$\Phi^{2a}$ in Eqs. (26) and (27), which is a canonical conjugate
with $\Phi^{1a}$,
\begin{eqnarray}
S_{\rm WZ} &=& \int d^3 x~ \left( \frac{\kappa}{2}
            ( (D_0\lambda)^a + \sqrt{\kappa}
               F^{20a} - \frac{1}{\sqrt \kappa} J^{1a})
              M^{-1}_{ab}
              ((D_0\lambda)^b + {\sqrt \kappa}
              F^{20b} - \frac{1}{\sqrt \kappa} J^{1b}) \right. \nonumber \\
        &-&  \left.  \frac{1}{2\kappa} \lambda^a M_{ab} \lambda^b
              + \lambda^a({\sqrt \kappa} F^{10a} +
              \frac{1}{\sqrt \kappa} J^{2a}) \right),
\end{eqnarray}
with the measure
\begin{equation}
[{\cal D} \mu] = {\cal D} A^{\mu a}
                {\cal D} \phi^{\dagger}
                {\cal D} \phi
                {\cal D}\lambda^{a}
               {\cal D}\xi^{a} \prod^3_{\beta = 0}
               \delta(G^a_{\alpha} [A^a_0 + \xi^a , A^a_i ,
               \lambda^{a}])
               \mid det \{\Gamma^a_{\alpha}, G^b_{\beta}\} \mid
               (det M_{ab})^{-1/2} ,
\end{equation}
where we redefine $\Phi^{1a}$ as $\lambda^a$,
and $M_{ab} = \phi^{\dagger} (T^aT^b + T^bT^a) \phi$.
Note that the local gauge symmetry of the Wess-Zumino action naturally also
survives in the configuration space. This suggests that the origin of
our second-class system is irrelevant to the conventional gauge-variant
Wess-Zumino like action [8,10,11]
which cancels the local gauge anomaly of the second-class system.
Interestingly just like the abelian case [13], the choice of $\lambda$=0
does not recover the original theory.

In conclusions, we have applied the Batalin-Tyutin Hamiltonian quantization
method to the nonabelian CS field theory as a good illustration to
show explicitly how to
convert the second-class system into first-class one
since it is special in that the gauge constraint algebra is weakly
first-class due to
the nonabelian group structure and the origin of the second-class algebra
is related to the symplectic structure irrelevant to the local gauge anomaly.
As a result, the gauge-invariant Wess-Zumino type action was obtained.
For further study, the role of the extended gauge symmetry in the extended
phase space (or configuration space) and the other
physical significance of the Wess-Zumino like action remain.

\section*{Acknowledgements}
We would like to thank Y.-W. Kim for discussions.
One of us (W.T. Kim) was supported in part by the Korea Science
and Engineering Foundation through the Center for Theoretical physics (1994).
The present study was also supported in part by
the Basic Science Research Institute Program,
Ministry of Education, 1993, Project No. 236.


\newpage

\begin{thebibliography}{99}

\bibitem{} P. A. M. Dirac, ``{\it Lectures on quantum mechanics}"
( Belfer graduate School, Yeshiba University Press, New York 1964 ).
\bibitem{} E. S. Fradkin and G. A. Vilkovisky, Phys. Lett. {\bf B55} (1975)
224.
\bibitem{} M. Henneaux, Phys. Rep. {\bf C126} (1985) 1.
\bibitem{} C. Becci, A. Rouet and R. Stora, Ann. Phys. [N.Y.] {\bf 98} (1976)
287;
I. V. Tyutin, Lebedev Preprint 39 (1975).
\bibitem{} T. Kugo and I. Ojima, Prog. Theor. Phys. Suppl. {\bf 66} (1979) 1.
\bibitem{} I. A. Batalin and E. S. Fradkin, Nucl. Phys. {\bf B279} (1987) 514;
    Phys. Lett. {\bf B180} (1986) 157.
\bibitem{} I. A. Batalin and I. V. Tyutin, Int. J. Mod. Phys. {\bf A6} (1991)
3255.
\bibitem{} L. D. Faddeev and S. L. Shatashivili, Phys. Lett. {\bf B167} (1986)
225;
O. Babelon, F. A. Shaposnik and C. M. Vialett, Phys. Lett. {\bf B177} (1986)
385;
K. Harada and I. Tsutsui, Phys. Lett. {\bf B183} (1987) 311.
\bibitem{} J. Wess and B. Zumino, Phys. Lett. {\bf B37} (1971) 95.
\bibitem{} T. Fujiwara, Y. Igarashi and J. Kubo, Nucl. Phys. {\bf B341} (1990)
695.
\bibitem{} Y.-W. Kim, S.-K. Kim, W. T. Kim, Y.-J. Park, K.Y. Kim, and Y. Kim,
Phys. Rev. {\bf D46} (1992) 4574;
R. Banerjee, H. J. Rothe and K. D. Rothe, Phys. Rev. {\bf D48} (1994) (in
press).
\bibitem{} T. Fujiwara, Y. Igarashi and J. Kubo, Phys. Lett. {\bf B251}
           (1990) 427; Y. Igarashi, H. Imai, S. Kitakado, J. Kubo, and H. So,
           Mod. Phys. Lett. {\bf A5} (1990) 1663.
\bibitem{} R. Banerjee, Phys. Rev. {\bf D48} (1993) R5467.
\bibitem{} R. Jackiw, ``{\it Topological Investigations of Quantized Gauge
Theories}",
edited by S. Treiman, R. Jackiw, B. Zumino and E. Witten
(World Scientific, Singapore 1985).
\bibitem{} G. Semenoff, Phys. Rev. Lett. {\bf 61} (1988) 517;
G. Semenoff and P. Sodano, Nucl. Phys. {\bf B328} (1989) 753.
\bibitem{} R. Banerjee, Phys. Rev. Lett. {\bf 69} (1992) 17;
        Phys. Rev. {\bf D48} (1993) 2905.
\bibitem{} N. Banerjee, S. Ghosh and R. Banerjee, Nucl. Phys. B (1994) (in
press);
Phys. Rev. {\bf D49} (1994) 1996.
\bibitem{} N. Banerjee, R. Banerjee and S. Ghosh,
``{\it Quantization of second-class systems in the Batalin-Tyutin formalism}",
Saha Institute Report, March 1994 ( hep-th 9403069 ).
\bibitem{} L. D. Faddeev and V. N. Popov, Phys. Lett. {\bf B25} (1967) 29.
\bibitem{} R. Banerjee, A. Chatterjee and V. V. Sreedhar,
Ann. Phys. (N.Y.) {\bf 122} (1993) 254.

\end{thebibliography}
\end{document}